\documentclass[aps,twocolumn,showpacs]{revtex4}
\usepackage[dvips]{graphicx}
\usepackage{amsmath}

\begin{document}

\title{Optical and Thermal Transport Properties of an
Inhomogeneous $d$-Wave Superconductor}
\author{W. A. Atkinson$^1$ and P. J. Hirschfeld$^2$}

\affiliation{
$^1$ Department of Physics, Southern Illinois
University, Carbondale, IL 62901-4401 \\
$^2$ Department of Physics, University of Florida, PO Box 118440,
Gainesville FL 32611 }
\date{\today}

\begin{abstract}
We calculate transport properties of disordered 2D $d$-wave
superconductors from solutions of the Bogoliubov-de Gennes
equations, and show that weak localization effects give rise to a
finite frequency peak in the optical conductivity similar to that
observed in experiments on disordered cuprates.  At low energies,
order parameter inhomogeneities induce linear and quadratic
temperature dependencies in microwave and thermal conductivities
respectively, and appear to drive the system towards a
quasiparticle insulating phase.

\end{abstract}

\maketitle

{\it Introduction.}
The study of disorder in high $T_c$ superconductors (HTSC) remains an
engaging topic for at least two reasons: first, it is apparent that
significant disorder---perhaps originating with charge-donor
impurities---is present in nearly all HTSC samples and, second, the
controlled doping of substitutional impurities is a powerful means of
studying the electronic state.  Transport experiments on the cuprates
at low $T$ have given us information on the quasiparticle lifetime in
the disordered superconductor and indicated the existence of strong,
near-unitarity limit scattering potentials associated with impurities
in the CuO$_2$ plane.  The simplest BCS-quasiparticle theories have
been successful at describing qualitative features of transport
experiments, but fail to explain many of the details.  This is the
principal motivation for the current work addressing transport
properties of dirty $d$-wave superconductors.  The approach taken here
is to model the paired state as an inhomogeneous superfluid via the
Bogoliubov-deGennes (BdG) equations.  We focus most of our attention
on optimally doped superconductors at low temperatures, where
inelastic processes freeze out\cite{freeze}
and mean-field theory is most applicable.

It is well known that HTSC are strongly affected by disorder because
of the $d$-wave symmetry of the pair order parameter $\Delta_{ij}$
($i$ and $j$ are site indices of the paired electrons).  In a pure
sample, the density of states (DOS) $\rho(\omega)$ is gapless and
vanishes as $|\omega|$ at the Fermi energy (taken to be 0 here).  A
single strong-scattering impurity produces a pair of subgap resonances
at $\pm \omega_0$ ($\omega_0 <\Delta_\mathit{max}$,
$\Delta_\mathit{max}$ the DOS peak associated with the gap edge in
tunneling experiments).  When a finite concentration $n_i$ of
impurities with impurity potential $U$ is present, the isolated
resonances are split, and broaden into an ``impurity band'' centered
at the Fermi energy.  The energy scale $\gamma$ of the impurity band,
below which $\rho(E)$ crosses over from linear ($|E| > \gamma$) to
constant ($|E| < \gamma$), is determined by $n_i$ and $U$.  These
essential features are captured in the widely-used self-consistent
$T$-matrix approximation (SCTMA) for impurity
scattering\cite{yashenkin}.  The SCTMA is a
perturbative scheme which is useful for treating point-like
scatterers.  It correctly incorporates physics associated with strong
scattering potentials, but ignores correlation effects between
impurities (i.e. localization effects), as well as the local response of the
superfluid to the impurities.  In the SCTMA, $\gamma$
is also the quasiparticle scattering
rate in the impurity band.

While the SCTMA-based notion of an impurity band appears to be fairly
consistent with the observed thermodynamic properties of the optimally
doped cuprates\cite{scalapino}, the simplest theory based on this
picture disagree with transport experiments.  Notably, the
low-temperature behavior of both the thermal and microwave
conductivities in several systems disagree with the simple prediction,
$\sigma,\kappa/T\sim T^2$ \cite{hirschfeld}. For example, the
low-temperature microwave conductivity in YBa$_2$Cu$_3$O$_{7-\delta}$
appears to vary roughly linearly with temperature, $\sigma \sim
T$\cite{hosseini}.  In addition, the optical conductivity of
disordered cuprates is observed to have a maximum at a
disorder-dependent frequency of order 100 cm$^{-1}$\cite{basov}; this
feature is also not found in the simple SCTMA analysis.\cite{LTpaper}
Finally, the SCTMA predicts the universality of residual transport
coefficients, i.e.  limiting values of $\kappa/T$ and $\sigma$ as
$T\rightarrow 0$ which depend only weakly on disorder\cite{lee}.
While this has been confirmed in thermal conductivity measurements on
YBa$_2$Cu$_3$O$_{7-\delta}$\cite{taillefer} and
Bi$_2$Sr$_2$CaCu$_2$O$_8$\cite{behnia}, there are other systems where
universality is not seen.\cite{otherzerokappa} We show below that some
of these discrepancies can be understood within a BCS framework by
going beyond the SCTMA.

{\it Approach.}  The BdG equations will be solved at two levels of
approximation.  Like the SCTMA, non-self-consistent (NSC) solutions
assume that $\Delta_{ij}$ is homogeneous, but unlike the SCTMA, NSC
solutions incorporate quantum coherence (ie.\ localization) effects
associated with scattering from multiple impurities exactly.
Self-consistent (SC) BdG solutions involve a further step in which the
nonlinear response of $\Delta_{ij}$ to the local disorder potential is
determined.  In both cases, the BdG equations are solved on a
tight-binding lattice with $N=1600$ sites and up to 50 disorder
configurations.  In matrix form, the mean-field Hamiltonian is
\begin{equation}
{\cal H}= \sum_{ij} \Phi_i^\dagger
\left [ \begin{array}{cc}
                 t_{ij} & \Delta_{ij} \\
    \Delta_{ij}^\dagger & -t_{ij}^\ast
\end{array} \right ] \Phi_j
\label{eq:ham}
\end{equation}
with $\Phi_i^\dagger = ( c_{i\uparrow}^\dagger, c_{i\downarrow} )$.
The subscripts $i$ and $j$ refer to site indices, and $t_{ij} = -t
\delta_{\langle i,j \rangle} + (U_i-\mu) \delta_{i,j}$ with
$\delta_{\langle i,j \rangle} = 1$ for nearest neighbour sites, and 0
otherwise.  All energies in this work are measured in units of $t$,
and the lattice constant is $a=1$.  The bond order-parameter is
$\Delta_{ij} = -V\langle c_{j\downarrow} c_{i\uparrow}\rangle$ with
$V$ the nearest neighbour pairing interaction.  The pure $d$-wave
superconducting state occurs in the disorder-free limit and is related
to the bond order parameters by $\Delta_{ij} = \frac{1}{2}\Delta_0
[(-1)^{x_{ij}} - (-1)^{y_{ij}}]$ with $(x_{ij},y_{ij})$ connecting
sites $i$ and $j$, and where $\Delta_0$ is the homogeneous $d$-wave
amplitude.  Spatial fluctuations arise naturally when one solves
$\Delta_{ij}$ self-consistently in the presence of a disorder
potential.  For this work $V=3.28$, making $\Delta_0=0.8$ which is a
factor of $\sim 4$ larger than the realistic case.  The eigenstates
are found using standard linear algebra routines to diagonalise
Eq.~\ref{eq:ham}.  The quasiparticle DOS is $\rho(\omega) =
N^{-1}\langle \sum_n \delta(\omega - E_n) \rangle$ where $E_n$ are the
eigenenergies for a given impurity configuration and
$\langle\ldots\rangle$ represents configuration averaging.

The complex conductivity
is
\begin{equation}
\sigma(\omega,T) = \left \langle
\frac{e^2 \hbar}{i\omega \pi N} \sum_{n,n^\prime}
|\hat \gamma^0_{nn^\prime} |^2 \frac{f(E_n)-f(E_{n^\prime})}
{\hbar\omega^+ -E_n+E_{n^\prime}}
\right \rangle,
\label{eq:cond}
\end{equation}
where $\hat \gamma^\alpha_{nm} \equiv \langle n| (p_x/m) \otimes
\tau^\alpha | m \rangle$ is the matrix element of the velocity between
eigenstates $n$ and $m$, $\tau^\alpha$ ($\alpha = 0,\ldots,3$) is
the Pauli matrix in particle-hole space, and $\omega^+ = \omega+i0^+$.
We use a binning procedure to evaluate the real part
of $\sigma(\omega,T)$.    Note the expression
(\ref{eq:cond}) is not manifestly gauge invariant, but we
do not expect this to be a problem since the collective response
of the one-component charged order parameter  occurs
at the plasma frequency.

\begin{figure}[tb]
\begin{center}
\leavevmode
\includegraphics[width=\columnwidth]{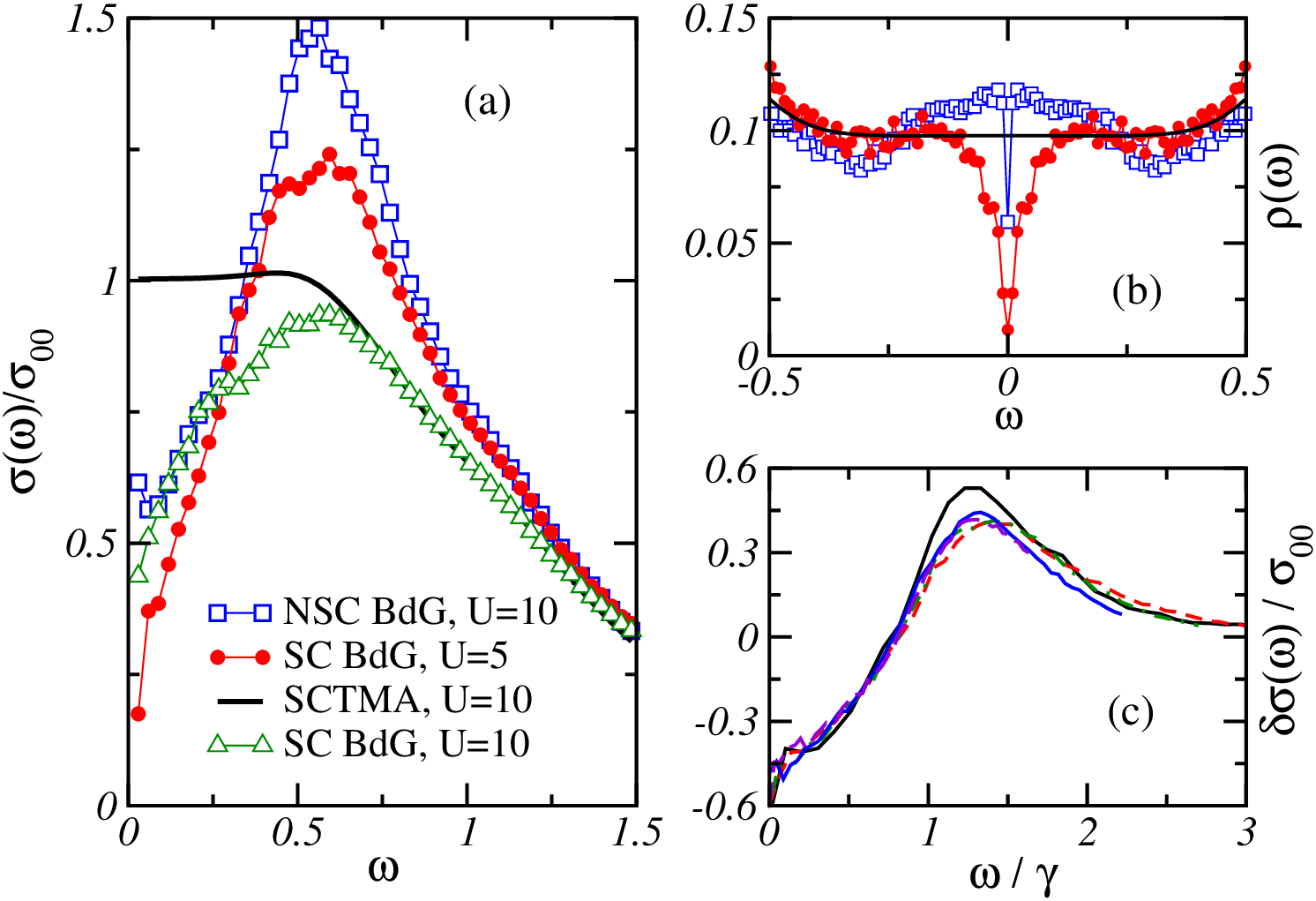}
\caption{Finite frequency conductivity for $\mu = 1.2$, $T=0$, and
$\Delta=0.8$ (ie.\ $v_F/v_\Delta=2.5$).  (a) Curves are for
self-consistently determined order parameter (SC BdG), homogeneous
order-parameter (NSC BdG) and SCTMA with $n_i=0.04$.  (b) Density of
states for NSC BdG and SCTMA with $U=10$, and SC BdG with $U=5$
(unitarity limit).  The gap-edge for the tunneling density of states
is $\Delta_{\mathit max} \approx 1.$,the SCTMA impurity bandwidth is
$\gamma \approx 0.45$. (c) Scaling of the NSC BdG conductivity for
$U=10$ and $n_i = 0.02$, 0.04, 0.06, 0.08, and 0.14.  Corresponding
$\gamma$ are $\gamma = 0.29$, 0.43, 0.55, 0.67, 0.91.  A small
deviation arises for $n_i = 0.02$ from finite size effects.  }
\label{fig:one}
\end{center}
\end{figure}

For numerical reasons, we are restricted to evaluating
$\sigma(\omega,T)$ in the ``gapless regime", $\gamma \agt T$,
which has been studied in intentionally damaged samples of
YBa$_2$Cu$_3$O$_{7-\delta}$\cite{zhang,basov}. In this regime, the
SCTMA predicts\cite{hirschfeld} that the real
part of the conductivity is $\sigma_{SCTMA}(\omega\rightarrow 0,T)
= \sigma_{00} + \alpha (T/\gamma)^2$ and $\sigma_{SCTMA}(\omega,0)
= \sigma_{00} + \alpha^\prime (\omega/\gamma)^2$, with a universal
value for the residual conductivity: $\sigma_{00} = e^2
v_F/(\pi^2\hbar v_\Delta)$.  In this expression, $v_F$ is the
Fermi velocity and $v_\Delta = |\nabla_k \Delta_k|$ is the
quasiparticle velocity component parallel to the Fermi surface.
Both impurity vertex corrections and Fermi-liquid corrections
renormalise $\sigma_{00}$ in an approximation\cite{durst} where
$s$-wave scattering is generalized to include anisotropic
components, but where weak localization corrections and order
parameter inhomogeneities are neglected.  In this scheme the
thermal conductivity is not renormalized\cite{durst}, $\kappa(T)/T
= \kappa_{00} + a(T/\gamma)^2$, with universal value $\kappa_{00}
= \frac{1}{3} k_B^2 (v_F^2 +v_\Delta^2)/v_Fv_\Delta$ which
survives this class of perturbative corrections.

{\it Results.}  Typical results for $\rho(\omega)$ near
unitarity---defined by $\omega_0 \ll \gamma$ and corresponding to
$U\approx 10$ and $U\approx 5$ in the NSC and SC calculations
respectively---are shown in Fig.~\ref{fig:one}.  NSC calculations
agree semiquantitatively with SCTMA calculations except below an
exponentially small energy\cite{atkinson1}.  The SC result, by
contrast, shows a large disorder-induced suppression of the DOS
relative to the SCTMA plateau.  As discussed
elsewhere\cite{atkinson1} the ``disorder-induced pseudogap'' (DIP)
at the Fermi-energy appears to be a generic feature of SC BdG
solutions and has an energy scale related to $\omega_0$ but which
grows with increasing $n_i$.  We stress that for typical planar Cu
substituents
in HTSC, this is an energy scale which is comparable to those
explored in transport experiments.

Figure~\ref{fig:one} shows the basic low-$T$ result for the
conductivity with strong scattering impurities.
$\sigma_{SCTMA}(\omega)$ is approximately Drude-like for $\omega >
\gamma$, but saturates at the universal value\cite{lee,LTpaper}
$\sigma_{00}$ for $\omega < \gamma$.  Numerical solutions of the
BdG equations deviate significantly from this with
$\sigma(\omega<\gamma)$ linear in frequency, and rising to a peak
at $\omega \approx \gamma$.  This is true for both $\sigma_{SC}$
and $\sigma_{NSC}$ (the SC and NSC BdG conductivities
respectively), and is therefore the result of weak localization
corrections to the SCTMA result\cite{larkin}. Indeed
Fig.~\ref{fig:one}(c) shows that $\delta \sigma(\omega) \equiv
\sigma_{NSC}(\omega) - \sigma_{SCTMA}(\omega)$ satisfies a scaling
relation, $\delta \sigma(\omega)/\sigma_{00} = F(\omega/\gamma)$,
which is similar to the weak-localization scaling relation for
dirty 2D metals.
In contrast, $\sigma_{SC}(\omega)$
does not display a simple scaling relation, as we discuss below.  At
this point, we simply remark that $\sigma_{SC}(\omega)$ always has
less finite-$\omega$ spectral weight than $\sigma_{SCTMA}(\omega)$ for
the same value of $U$ [as illustrated in Fig.~\ref{fig:one}(a)], with
the lost weight appearing in the superfluid response\cite{randeria}.

The finite-frequency conductivity peak exhibited in Fig.~\ref{fig:one} is
reminiscent of conductivity measurements in disordered HTSC\cite{basov}.
Experimentally, the peak is also seen in the normal state $T>T_c$, generally at
higher energies, whereas in our approximation it occurs at significantly lower
energies, and is much less pronounced than in the superconducting state.  It is
clear that inelastic scattering is important in the normal state and needs to
be incorporated in a complete explanation of the finite-frequency peak.
Nevertheless, this is the simplest way of understanding
this feature of the optical data, which has not been reproduced in any other
approach to our knowledge.

\begin{figure}[b]
\begin{center}
\leavevmode
\includegraphics[width=\columnwidth]{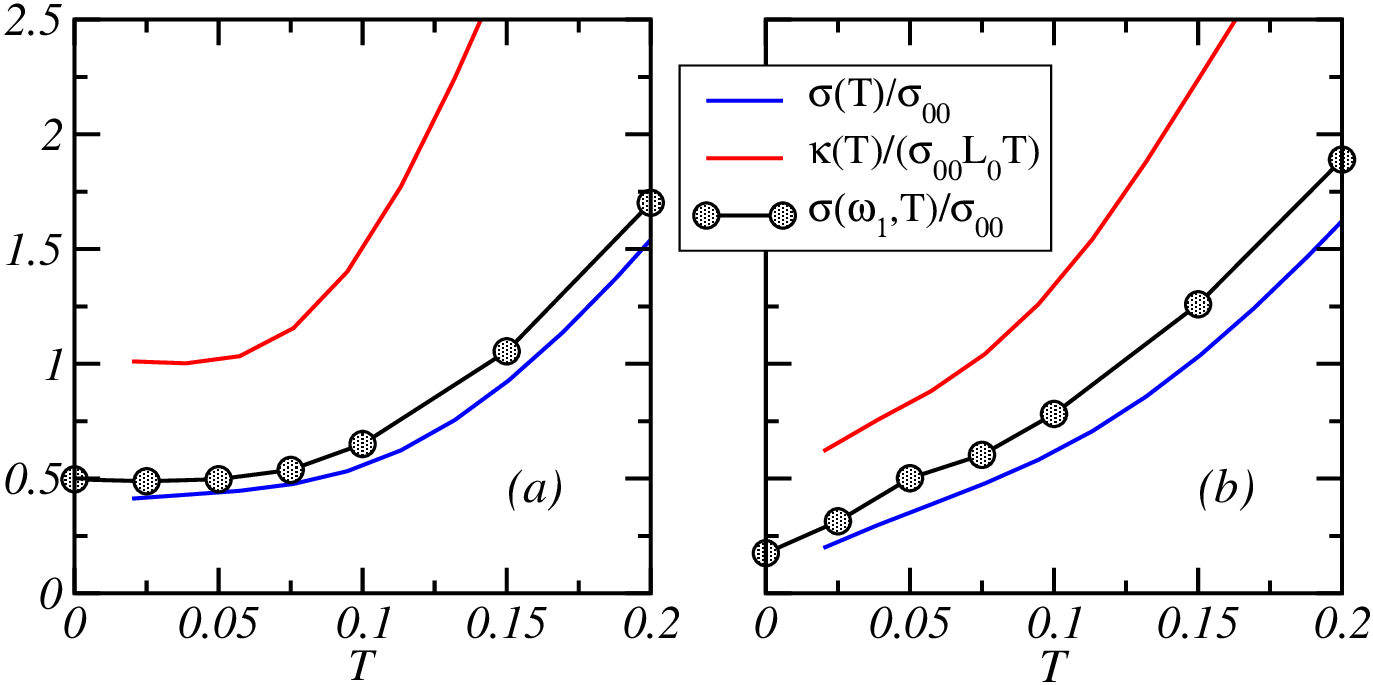}
\caption{$T$-dependent conductivity, normalised to $\sigma_{00}$ for
(a) NSC BdG and (b) SC BdG with $n_i=0.06$, $U=10$, $\mu = 1.2$, and
$\Delta=0.8$.  $\omega_1=0.0297$ is the lowest nonzero frequency used
in calculating $\sigma$ with Eq.\ (\protect\ref{eq:cond});
$\kappa/(TL_0)$ $\sigma(T)$ are evaluated using Eqs.~(\protect\ref{4})
and (\protect\ref{3}).  }
\label{fig:two}
\end{center}
\end{figure}

The temperature dependence of the low-frequency conductivity is
also of experimental interest.  In Fig.~\ref{fig:two} we have
plotted $\sigma(\omega_1,T)$, where $\omega_1 = 0.0297$ is the
lowest nonzero frequency, chosen since $\omega=0$ suffers from
finite-size effects. The strong $T$-dependence of
$\sigma_{NSC}(\omega_1,T)$ is similar to the SCTMA result.  On the
other hand, $\sigma_{SC}(\omega_1,T)$ has a linear-$T$
conductivity, reminiscent of most microwave conductivity
experiments\cite{hosseini}.  These results are generic for a wide
range of $U$ near unitarity. Figure \ref{fig:two} also shows the
thermal conductivity
\begin{equation}
\kappa(T) = \frac{1}{2\pi \hbar T} \int dx\, x^2\, \left( -\frac{\partial
f}{\partial x} \right ) \left \langle S_T(x) \right \rangle,
\label{e3a}
\end{equation}
where
\begin{equation}
S_T(x) = \frac{2\pi^2\hbar^2}{N} \sum_{n,n^\prime}
|\langle \hat v_g \rangle_{nn^\prime} |^2
\delta(x-E_n) \delta(E_n-E_{n^\prime}),
\label{4}
\end{equation}
with quasiparticle group velocity $\hat v_g = \hat \gamma^3 + {\hat
v_\Delta}^x \otimes \tau^1$, and $\langle \hat v_\Delta^x\rangle_{ij}
= (i/\hbar)(x_i-x_j)\Delta_{ij}$ in the site
representation\cite{kampf}.  For a finite system, the
$\delta$-functions in Eq.~(\ref{4}) are broadened to smooth the
discreteness of the energy spectrum.  It is apparent in Fig.\
\ref{fig:two} that the Wiedemann-Franz law $\kappa/\sigma T = L_0
\equiv k_B^2\pi^2/3e^2$, which is already violated because of
differences between the group and Fermi velocity, appears to also be
violated at low $T$ (at least in the NSC case) by weak localization
corrections.  For comparison, we show a similar calculation of the
charge-conductivity which becomes exact in the limit $\omega \ll T$,
\begin{equation}
\sigma(T) = \frac{e^2}{2\pi\hbar} \int dx \,
\left ( -\frac{\partial f(x)}{\partial x} \right )
\left \langle S_\sigma(x) \right \rangle
\label{3}
\end{equation}
with $S_\sigma$ identical to $S_T$ with the replacement of $\hat
v_g$ by $\hat \gamma^0$.  From the figure, it is clear that Eq.\
(\ref{3}) is in good agreement with $\sigma(\omega_1,T)$, and that
both $\kappa(T)/T$ and $\sigma(T)$ exhibit a linear $T$-dependence
over a wide range of temperatures.
The extent of the linear regime is discussed below.
Linear power laws have been claimed in thermal conductivity
measurements\cite{behnia} (but other power laws have also been
reported), and this work provides a potential mechanism.
Conductivities with odd power-laws in $T$
are difficult to achieve in the SCTMA because all quantities in the
gapless regime are analytic functions of $\omega$.  We note that
quasilinear behavior over some intermediate temperature regime has
been obtained in SCTMA approaches, by invoking a special combination
of unitarity and weak scatterers\cite{berlinsky}, by fine tuning
scattering phase-shifts\cite{scharnberg}, or by including order
parameter fluctuations in a SCTMA-like approximation\cite{hettler}.

\begin{figure}[b]
\begin{center}
\leavevmode
\includegraphics[width=\columnwidth]{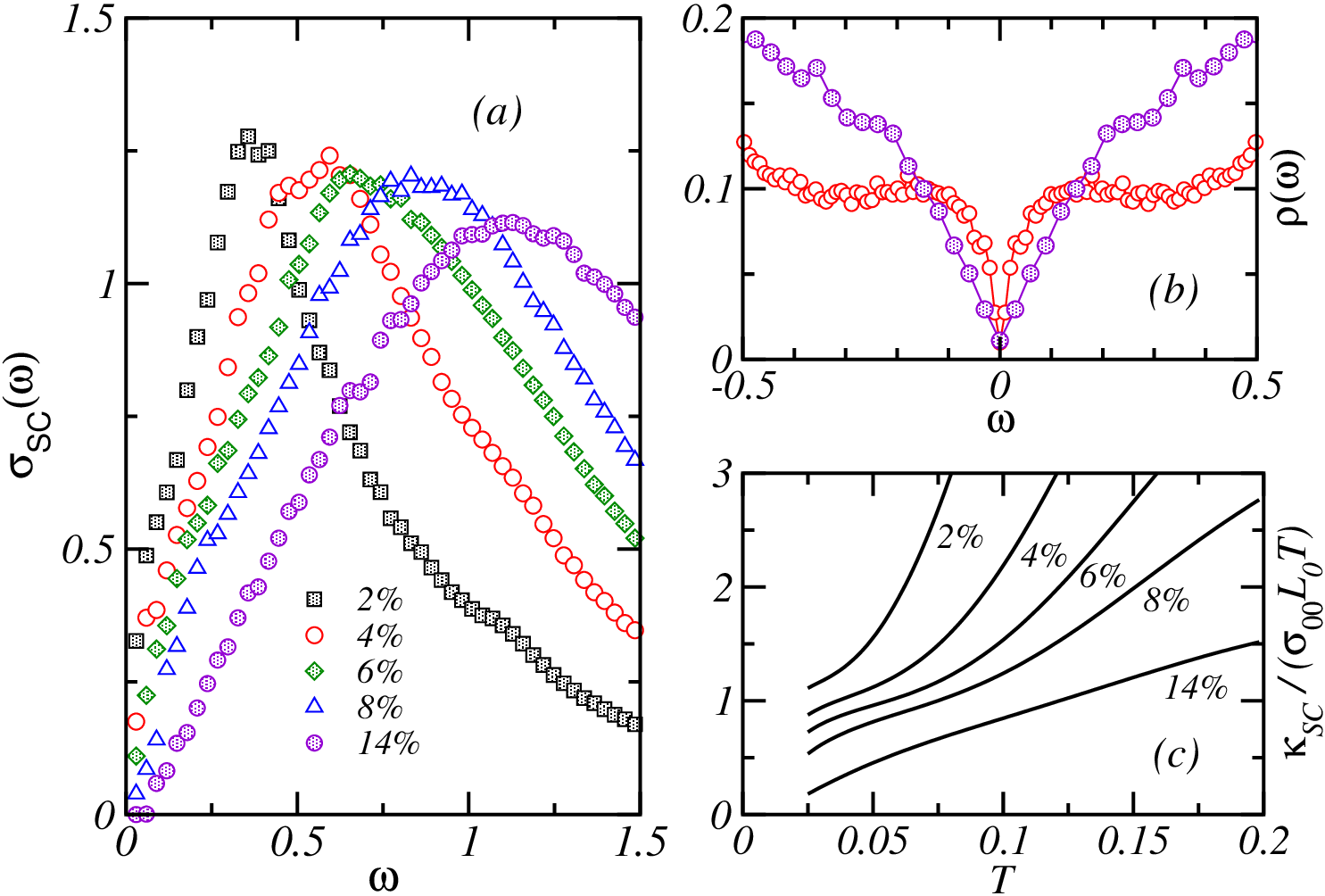}
\caption{Scaling of $\sigma_{SC}(\omega)$ with $n_i$ at $T=0$.  (a)
$\sigma_{SC}(\omega)$ for a range of $n_i$ between 0.02 and 0.14,
$U=5$.  (b) Density of states at $n_i = 0.04$ (open circles) and $n_i
= 0.14$ (filled circles).  (c) Thermal conductivity vs. temperature
for same $n_i,U$.  } 
\label{fig:three}
\end{center}
\end{figure}
Finally, in Fig.\ \ref{fig:three} we study the dependence of
$\sigma_{SC}(\omega)$ on impurity concentration.  As $n_i$ is
increased, the peak position in $\sigma_{SC}$ increases, in
qualitative agreement with the scaling of $\sigma_{NSC}(\omega)$.  The
scaling of $\sigma_{SC}(\omega)$ is not straightforward, however,
since low-frequency spectral weight is depleted as $n_i$ increases, in
contrast to $\sigma_{NSC}(\omega)$ which depends only weakly on $n_i$
at low $\omega$.  The depletion is correlated with the growth of the
DIP, shown in Fig.~\ref{fig:three}(b), reminiscent of disordered
interacting metals near the metal-insulator transition.  For the model
interaction chosen, we never observe a transition to a truly gapped
state, and it is doubtful that a quasiparticle metal-insulator
transition could be observed in real HTSC since superconductivity is
destroyed in heavily damaged samples.  However, the current work is
strongly suggestive that the physics of disordered cuprate
superconductors is influenced by proximity to such a transition.

In Fig.\ \ref{fig:three}(c), we plot the SC thermal conductivity for
several impurity concentrations to illustrate the robustness of the
(quasi)linear-$T$ regime. The regime is bounded by two
disorder-dependent temperature scales; the upper crossover temperature
is readily apparent up to $n_i = 0.08$, and is correlated with the
weak localization scale $\gamma$, while the lower bound signals a
downturn in $\kappa/T$ which appears to scale with the DIP.  For $n_i
= 0.14$, there is no clear distinction between these scales. With
current system sizes it is difficult to determine the lowest energy
behavior. If one assumes the matrix elements of $\langle \hat v_g
\rangle$ have only weak energy dependence near the Fermi surface, then
we expect $S_T(x) \sim \rho(x)^2$ which is $\sim x^{2\alpha}$ near
$x=0$ ($\rho \sim x^\alpha$\cite{atkinson1}) in the SC BdG
calculations.  For sufficiently small $T$, then, one anticipates a
downturn with $\kappa/T \sim T^{2\alpha}$ below the DIP energy
scale. It is clear that a strong suppression relative to the universal
SCTMA result $\kappa(T)\rightarrow \kappa_{00}$ is to be expected.

 {\it Conclusions.}  We have observed effects with two distinct
physical origins in this work.  First, there is a pronounced peak in
$\sigma(\omega)$ at $\omega\approx \gamma$ arising from localization
physics.  This occurs whether or not the BdG equations are solved
self-consistently, and is consistent with the experimental fact that
the peak is only observed in very disordered systems.  Since $\gamma
\sim \sqrt{n_i}$ for strong scatterers, our work suggests that a
systematic study of samples with varying impurity concentrations will
provide an experimental means to distinguish between the weak
localization mechanism presented here, and other proposed origins for
the peak.  Second, we have found that important physics associated
with the correlated order-parameter response to disorder arises at low
energies. Perhaps the most striking result is the observation of
linear-$T$ power laws in the charge and thermal conductivities, as
observed in some high-$T_c$ systems.  In addition, order parameter
supression effects appear to eliminate the residual conductivities at
asymptotically low temperatures expected on the basis of SCTMA and
other treatments.  We note that the current work assumes fairly
disordered systems, and the extrapolation to the clean limit ($\gamma
< T$) is not obvious.  On the other hand, this is the first time that
this additional source of off-diagonal scattering has been correctly
accounted for in a transport theory, and there appears to be no reason
in principle why these effects should also not be important in clean
2D systems and possibly even in higher dimensions. To compare directly
with experiments, the effect of realistic Dirac cone anisotropies and
inelastic scattering needs to be better understood.  Work along these
lines is in progress.

{\it Acknowledgements.} This work is supported by NSF grant DMR-9974396.
The authors would like to thank A. Kampf and N. Trivedi for  useful
discussions.


\begin{thebibliography}{yashenkin}
\bibitem{kampf} Note that we have simply discretized the group velocity
operator which gives the thermal current in the uniform case.  Additional
terms arise when deriving the thermal current in an inhomogeneous system
which we have neglected here. A. Kampf (private communication).
\end{thebibliography}
\end{document}